# Voltage Control Using Eigen Value Decomposition of Fast Decoupled Load Flow Jacobian

Ali Dehghan Banadaki[1], *Student Member, IEEE*, Ali Feliachi[1], *Senior Member, IEEE*
Lane Department of Computer Science and Electrical Engineering,
West Virginia University, Morgantown, WV 26505, USA

*Abstract*— Voltage deviations occur frequently in power systems. If the violation at some buses falls outside the prescribed range, it will be necessary to correct the problem by controlling reactive power resources. In this paper, an optimal algorithm is proposed to solve this problem by identifying the voltage buses, that will have a maximum effect on the affected buses, and setting their new set-points. This algorithm is based on the Eigen-Value Decomposition of the fast decoupled load flow Jacobian matrix. Different Case studies including IEEE 9, 14, 30 and 57 bus systems have been used to verify the method.

I. INTRODUCTION:

Voltage control has been known as one of the important problems in power systems for decades [1- 4]. As an example, the blackout in Western Europe was caused by the phenomenon of voltage collapse [5]. Different voltage control devices for injecting reactive power have been used such as synchronous condensers [6], tap changer transformers [7], capacitor banks [8, 9], and Flexible AC transmission systems (FACTS) [10-16]. Voltage controllers are typically categorized in a hierarchical structure including three levels, namely primary, secondary and tertiary levels [17-21]. In the primary control, the voltage is controlled locally and instantly through the Automatic Voltage Control (AVR) [22]. It might be seen that the voltage profile is not still optimal although the primary control has been applied. Therefore, the secondary voltage controllers have been used to bring back the voltage to its nominal value. The time constant of the secondary controllers are more than the primary controller to prevent interaction between them. In the tertiary control, the control objectives such as economic dispatch will be met [23-27]. In this paper we are working on a secondary voltage controller (SVC).

In SVC, due to the local nature of voltage control and reactive power [28], regional analysis should be done; otherwise controlling a voltage bus from a long distance will need many efforts and might not be reasonable. Different papers proposed on SVC used the pilot buses in the system in which total voltage profile can be controlled by a class of disturbance [1, 29, 30]. The procedure of finding pilot buses is nontrivial since it can be a nonlinear and combinatorial problem [31]. By assuming different assumptions such as minimizing active power losses, reactive power injection or voltage deviation, they can reach different pilot buses [32]. In [19], the zoning evaluation has been done for AVR in the primary controller. Reference [33], used a sensitivity matrix to identify voltage control areas rather than proposing an algorithm to control the system. In [34], the voltage control has been proposed by considering the zonal effect but it has a high chance of trial and error [35].

An optimal control that is controlling the voltages of a power system with considering the zonal effect at the same time is exploited in our work in [36] . Since in transmission level, the reactance of the lines is much more than the resistance of the lines, reactive power control can be assumed to be decoupled from the active power. Therefore, fast decoupled load flow (FDLP) has been used to find system zonal areas. In other words, from FDLP, the sensitivity matrix is made that shows the sensitivity of load buses to generator buses. The Eigen Value Decomposition (EVD) on this sensitivity matrix has been applied to find the best set of inputs (voltage of generators) that can control the optimal output (voltage of violated buses). This set of inputs will indicate the participation of each generator to compensate a voltage violation. However, a conflict can happen in case of controlling the multiple voltage violations. In this paper, we clear the conflict by modifying the given set of inputs coming from the EVD based on the conflicted zone.

The remainder of this paper has been organized as follows: In Section II, the preliminary math on eigenvector and eigen-decomposition has been explained. In section III, Problem formulation and objectives are explained. In section IV, the proposed method has been implemented in four different IEEE case studies. Finally, in section V, the conclusion is given.

II. PRELIMINARY MATH ON EIGEN DECOMPOSITION:

An eigenvector of a matrix A is a special vector that when multiplied by matrix A, the result is another vector in the same direction of itself. However, the magnitude of the given vector will be scaled by its corresponding eigenvalue as it is shown in equation (1) and illustrated in figure (1):

$$AV = \Lambda V \quad (1)$$

Where $V = [v_1 v_2 ... v_n]$ is the set of eigenvectors of matrix A and $\Lambda = diag[\lambda_1 \lambda_2 ... \lambda_n]$ is the diagonal form of corresponding eigenvalues. This characteristic has been shown in the figure (1) for one eigenvector.

Given a real symmetric matrix A, all of the eigenvectors are mutually orthogonal for distinct eigenvalues. In other words, this means that each of the eigenvectors points to a direction that is perpendicular to the other one. For instance, eigenvectors of an arbitrary identity matrix N of size 2 are $v1 = [1 \quad 0]^T$ and $v2 = [0 \quad 1]^T$ that are orthogonal vectors. Now let us consider the set of linear equations in (2):

$$Y = AX \quad (2)$$

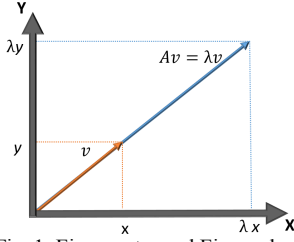

Fig. 1: Eigenvector and Eigenvalue

where A is a real matrix. A performance index is defined as the weighted L2-norm of the output vector Y as in equation (3):

$$J = Y^T M Y = X^T N X \quad (3)$$

where $N = A^T A$. Now we can represent matrix N with its eigenvectors and eigenvalues (i.e. eigen decomposition of matrix N) as in (4):

$$N = V . \Lambda . V^{-1} \quad (4)$$

It can be shown that when input X is chosen to be equal to one of the eigenvectors of matrix N in equation 3, the performance index J would be equal to its corresponding eigenvalue. (Proof of it has been given in the appendix for interested readers):

$$X = v_i \to J = X^T N X = v_i^T N v_i = \lambda_i \quad (5)$$

It is shown in figure (2) that by choosing each eigenvectors as an input, we would get different vectors that are in the direction of its own eigenvector (which is perpendicular to the other one) but with different magnitude that are related to its own eigenvalue. Therefore, if the objective is to find the maximum of J, we could easily choose the eigenvector corresponding to the largest eigenvalue of matrix N which we name it here $v_{\lambda_{max}}$ and it is shown in equation (6):

$$X = v_{\lambda_{max}} \to J_{max} = v_{\lambda_{max}}^T N v_{\lambda_{max}} = \lambda_{max} \quad (6)$$

It is worth mentioning that if N is not a square matrix, in a more general way, the same approach can be kept by choosing $X = U_1$, where $U_1$ is the left singular eigenvalue of the matrix $N$. In this case the maximum of $J$ is equal to:

$$J_{max} = X^T N X = \sigma_1^2 \quad (7)$$

Where $\sigma_1$ is the singular value of matrix $N$.

### III. PROBLEM FORMULATION:

#### A. Sensitivity Matrix:

Using the FDLP algorithm, the relationships between voltages of load buses and generator buses can be written by:

$$\begin{bmatrix} \Delta V_G \\ \Delta V_D \end{bmatrix} = \begin{bmatrix} S_{11} & S_{12} \\ S_{21} & S_{22} \end{bmatrix} \begin{bmatrix} \Delta Q_G \\ \Delta Q_D \end{bmatrix} \quad (8)$$

Where $\Delta V_G$ and $\Delta Q_G$ are respectively voltages and reactive power changes at generating buses while $\Delta V_D$ and $\Delta Q_D$ are respectively voltages and reactive power at load buses.

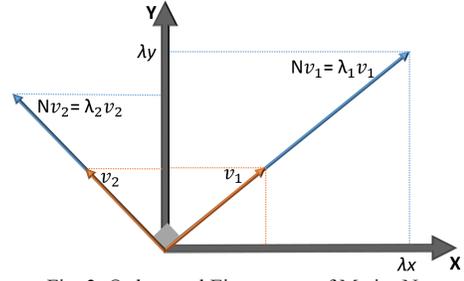

Fig. 2: Orthogonal Eigenvetors of Matirx N

Solving equation (8) for $\Delta V_D$ will lead to the equation (9):

$$\Delta V_D = S_{21} S_{11}^{-1} \Delta V_G + W \quad (9)$$

Where $W = (S_{22} - S_{21} S_{11}^{-1} S_{12}) \Delta Q_D$

In this method, $W$ is considered as a disturbance since reactive power changes in the load is not assumed to be available. On the other hand, the first term in equation (9) shows the effect of change in source voltages on the load voltages. This matrix is called sensitivity of voltage sources on loads and can be calculated in (10):

$$S_{GL} = S_{21} S_{11}^{-1} \quad (10)$$

For example, in the IEEE 9 bus system that has been shown in figure (4), from $S_{GL}$ matrix it can be deduced that bus 8 is more sensitive to generator 2 and then generator 3 and finally generator 1. This means that in order to control this bus voltage optimally, generator 2 should be used more than generator 3 and 1 until it reaches its maximum limit. Therefore, this matrix is a vital part of our effort to find the objective function in the next part.

#### B. The objective of the controller:

The objective of the controller is to find the best set of inputs $X = \Delta V_G$ that have the maximum effect on the outputs $Y = \Delta V_D$. The L2-norm has been used here as the performance index of changes in the output:

$$\text{Objective} \quad \max_{\Delta V_G} J = \Delta V_D^T M \Delta V_D \quad (11)$$
$$\text{s.t.} \quad |\Delta V_G| \leq \Delta V_{Gmax}$$

Where M is a square matrix with the size of load buses. Now by using equation (9) (neglecting the disturbance $W$), equation (11) can be rewritten as follows:

$$\max_{\Delta V_G} J = \Delta V_D^T M \Delta V_D \approx \Delta V_G^T N \Delta V_G \quad (12)$$

Where $N = S_{11}^{-T} S_{21}^T M S_{21} S_{11}^{-1}$

Based on the equation (6), this control objective can be solved by choosing the input as $\Delta V_G = \alpha . v_{\lambda_{max}}$, where $\alpha$ is a scaler that has been added as a design parameter and $v_{\lambda_{max}}$ is the eigenvector corresponding to the highest eigenvalue of the matrix $N$. To control a single bus $i$, matrix M is a zero matrix other than the entry of $(i,i)$ that is equal to 1. In this case, we can find $\alpha$ from equation (13) if no constraint is included:

$$J_{max} = \Delta V_{ctr}^T M \Delta V_{ctr} = \Delta V_{ctr}^2 = \alpha^2. \lambda_{max} \quad (13)$$

Therefore, by selecting desired $\Delta V_{ctr}$ which is the difference of controlled bus from 1 pu, $\alpha$ can be calculated. By using the calculated $\alpha$, the voltage of the control bus will move towards 1 pu but it might impose an extra effort on generators to meet this criteria which means constraint of equation (11) can been violated. Therefore, control input will be normalized in (14) so that the added control input plus the previous voltage of generators are not more than the maximum limit:

$$\Delta V_G = \alpha. v_{\lambda_{max}} \xrightarrow{|\Delta V_G| \nleq \Delta V_{G_{max}}} |\Delta V_G|_{norm} = \bar{\alpha}. v_{\lambda_{max}} \quad (14)$$

Where $\bar{\alpha} = \alpha. \frac{\Delta V_{G_{max}}}{\Delta V_G}$

In this approach controlling a single violated bus with considering the constraint on the generating voltages has been given. The same approach can be applied in the cases that multiple violated buses have happened that are either less than the lower limit (0.9 pu) or above the upper limit (1.1 pu). This process is called single-sided control procedure and the less violated bus is chosen to be the control bus (i.e. the one that is out of the limit but is closer to 1 pu among the other violated buses).

On the other hand, in the case of multiple violated buses where some of the voltages are above the upper limit and others are less than the lower limit, one control bus is chosen for each side with the same approach explained in the single-sided procedure. This is called the double-sided procedure and we need to check if any conflict of interest in the system exists. A conflict of interest is happened when controlling a lower limit violated bus (i.e. buses with the voltage less than 0.9 pu) will deteriorate the voltage of an upper limit voltage bus (i.e. buses with the voltage greater than 1.1 pu). In this case, control input should be updated by equation (15) to check if a conflict of interest exists in controlling bus voltages i and j at the same time,

$$V_{ctr} = \Delta V_{Gi} \cap \Delta V_{Gj} \quad (15)$$

Where symbol ∩ is used here as an operator that removes the confliction among the conflicted buses as in equation (16):

$$X \cap Y = \begin{cases} 0 & sign(x_i. y_i) < 0 \\ x_i + y_i & otherwise \end{cases} \quad (16)$$

For example, in figure (3) if the voltage of bus 1 is over 1.1 pu and the voltage of bus 2 is less than .9 pu, then controlling this system with the help of generator 1 would face a conflict. In figure (3.a) there is no way to control the voltages of both buses at the same time with the help of the generator one since increasing/decreasing the voltage of generator 1 will deteriorate the voltage of the other bus. However, in figure (3.b) it is possible to control the voltages of bus 1 and 2 by using generators 2 and 3 and neglecting generator 1 even if it is more sensitive to the affected buses in terms of the sensitivity matrix. In the following the proposed algorithm is explained:

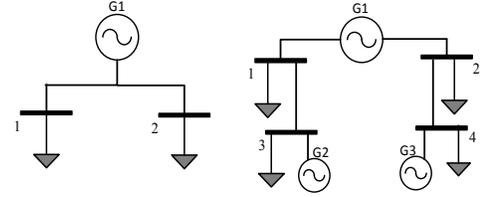

Fig. 3: An Example for Two sided-conflict

### C. Proposed Algorithm for optimal voltage control:

1- Solve the load flow and determine the violated bus(es).
2- End if there is no violated bus.
3- Determine the control bus in a single sided procedure or two control buses in a double sided procedure.
4- Determine $v_{\lambda_{max}}$ which is the eigenvector of the matrix $N$ in (12), corresponding to its highest eigenvalue.
5- Calculate the deviation: $\Delta V_{ctr} = 1 - V_{ctr}$.
6- Determine $\alpha$ from $\alpha^2 = \Delta V_{ctr}^2 / \lambda_{max}$.
7- Determine the control input $\Delta V_G = \alpha. v_{\lambda_{max}}$.
8- Update $\alpha$ by equation (14) if the generator voltages are not within their limits in step 4.
9- Control input should be checked and updated by equation (15) for double-sided conflict.
10- Update voltage of generators: $V_{Gnew} = V_{Gold} + |\Delta V_G|_{norm}$.
11- Go to Step 1.

### IV. CASE studies:

In this paper, four IEEE cases have been chosen to illustrate our method. Each case is initially in its normal mode (i.e. all voltages are within their normal limits). In order to show that the algorithm can control the voltages of the system, loads of type inductive or capacitive have been added to the original system to make some voltages go out of their limits. These simulations have been done in Matlab and load flow solution has been verified by Matpower package [37].

### A. Cae Study One: IEEE-9 bus system:

IEEE 9 bus system is shown in figure (4). It has 3 generating units which are producing 319.95 MW and 34.88 MVAr altogether. The total loads of the system are 315.00 MW and 115.00 MVAr. All the voltages are within their normal limits (i.e. .9 pu to 1.1 pu). 115 MVAr has been added to bus 7 to make one of the voltages go out of its limit. The algorithm has been applied to control any bus voltage that is out of the limit (here it is bus number 7). Control command for this system in order to bring back voltage of bus 7 is given here:
$[\Delta V_{G1} \quad \Delta V_{G2} \quad \Delta V_{G3}] = [0.0307 \quad 0.1 \quad 0.0851]$

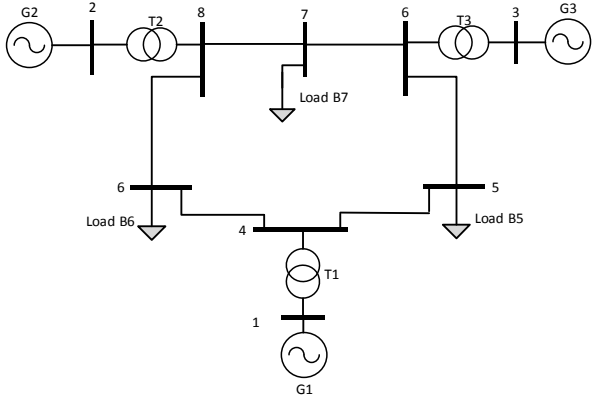

Fig. 4: IEEE 9 bus system

From this control input, it is obvious to see that the generators that are more sensitive to the controlled bus are participating more than the others.

TABLE I. Bus Voltages before and after Control
(*IEEE* – 9bus)

| Bus | Voltage (pu) | | |
|---|---|---|---|
| | *Normal* | *Disturbance* | *Control* |
| 1 | 1 | 1 | 1.031 |
| 2 | 1 | 1 | 1.1 |
| 3 | 1 | 1 | 1.085 |
| 4 | 0.987 | 0.974 | 1.029 |
| 5 | 0.975 | 0.955 | 1.025 |
| 6 | 1.003 | 0.971 | 1.06 |
| 7 | 0.986 | 0.891 | 0.994 |
| 8 | 0.996 | 0.955 | 1.053 |
| 9 | 0.958 | 0.933 | 1.008 |

TABLE II. Bus Voltages before and after Control
(*IEEE* – 9bus)

| Bus | Voltage (pu) | | |
|---|---|---|---|
| | *Normal* | *Disturbance* | *Control* |
| 1 | 1 | 1 | 1.1 |
| 2 | 1 | 1 | 1.018 |
| 3 | 1 | 1 | 1.018 |
| 4 | 0.987 | 0.883 | 0.977 |
| 5 | 0.975 | 0.84 | 0.922 |
| 6 | 1.003 | 0.97 | 1.006 |
| 7 | 0.986 | 0.956 | 0.993 |
| 8 | 0.996 | 0.971 | 1.007 |
| 9 | 0.958 | 0.876 | 0.955 |

In table I, voltages of the normal case (without applying any disturbance), disturbance case (before applying control) and control case (after applying control) is shown.

In another example, we might have a case that multiple buses are out of limit. To illustrate this case, a load of 150 MVAr has been added to bus number 4 and another load of 70 MVAr has been added to bus number 5. Three voltages of buses 4, 5 and 9 are going out of their limits. Algorithm will choose bus number 4 as a control bus since it has less deviation among the other violated buses. Results have been shown in table II.

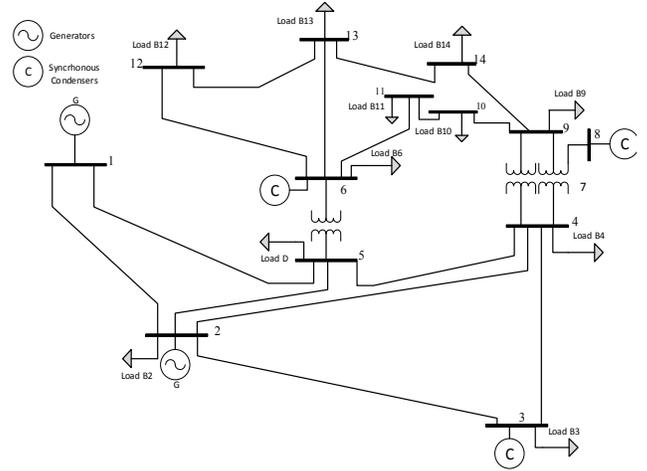

Fig. 5: IEEE 14 bus system

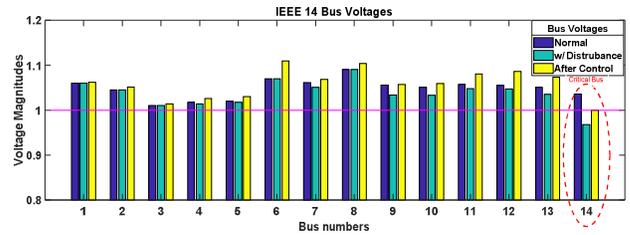

Fig. 6: Voltage Profile of IEEE 14 bus system [36]

TABLE III. Bus Voltages before and after Control
(*IEEE-14 bus*)

| Bus | Voltage (pu) | | |
|---|---|---|---|
| | *Normal* | *Disturbance* | *Control* |
| 1 | 1.06 | 1.06 | 1.09 |
| 2 | 1.045 | 1.045 | **1.1** |
| 3 | 1.01 | 1.01 | 1.03 |
| 4 | 1.018 | 0.929 | 0.97 |
| 5 | 1.02 | 0.874 | 0.92 |
| 6 | 1.07 | 1.07 | **1.1** |
| 7 | 1.062 | 1.021 | 1.049 |
| 8 | 1.09 | 1.09 | **1.1** |
| 9 | 1.056 | 1.016 | 1.05 |
| 10 | 1.051 | 1.018 | 1.05 |
| 11 | 1.057 | 1.04 | 1.07 |
| 12 | 1.055 | 1.052 | 1.09 |
| 13 | 1.05 | 1.045 | 1.08 |
| 14 | 1.036 | 1.01 | 1.04 |

*B. Case Study Two: IEEE-14 bus system:*

IEEE 14 bus system is shown in figure (5). It has 11 loads that consume 259 MW and 73.5 MVAr totally. A disturbance of 30 kVAR has been added to bus 14. The result of the algorithm after applying the control input to the disturbance case is shown in figure (6).

In another scenario, we have added 300 MVAr to bus 5. Since the amount of disturbance is too high, three generators reached their maximum limits to control the violated bus; Therefore, the algorithm brings back the voltage of bus 5 to 0.92 pu instead of 1 pu. The result has been shown in table III.

## C. Case Study Three : IEEE -30 bus system:

IEEE 30-bus system has 6 generators located at buses 1, 2, 13, 22, 23 and 27 . The total generating power are 191.64 MW and 100.41 MVAr. In this case study, we have considered multiple bus violations as well as having two-sided conflict in our system. For this example, we have added 130 MVAr, 40 MVAr, 40 MVAr, -35 MVAr and -35 MVAr to buses 28, 24, 19, 29 and 30 respectively. Voltage of buses 29 and 30 go above 1 pu while other voltages are lower than .9 pu. In this case, the first iteration has been done by working on bus 20 and bus 29 as a two-sided conflict. This will make bus 28 to go out of the limit. Therefore, we repeat the process for the second iteration by choosing bus 28 as the control bus which makes bus 22 to go out of its limit. Finally, by controlling bus 22 for the third iteration, all of the voltages go back to their normal limits.

TABLE IV. Bus voltages in each iteration
(*IEEE* – 30bus)

| Bus | Voltage (pu) | | | | Bus | Voltage (pu) | | | |
|---|---|---|---|---|---|---|---|---|---|
| | *Dist.* | *Iter. 1* | *Iter. 2* | *Iter.3* | | *Dist.* | *Iter. 1* | *Iter. 2* | *Iter.3* |
| 1 | 1.00 | 1.01 | 1.03 | 1.03 | 16 | 0.95 | 1.03 | 1.05 | 1.00 |
| 2 | 1.00 | 1.03 | 1.10 | 1.10 | 17 | 0.95 | 1.03 | 1.06 | 0.98 |
| 3 | 0.95 | 0.98 | 1.02 | 1.00 | 18 | 0.894 | 0.98 | 1.00 | 0.95 |
| 4 | 0.93 | 0.97 | 1.02 | 1.00 | 19 | 0.86 | 0.95 | 0.98 | 0.91 |
| 5 | 0.95 | 0.98 | 1.04 | 1.03 | 20 | 0.897 | 0.97 | 1.00 | 0.93 |
| 6 | 0.91 | 0.94 | 0.99 | 0.97 | 21 | 0.99 | 1.09 | 1.11 | 1.00 |
| 7 | 0.91 | 0.95 | 1.00 | 0.98 | 22 | 1.00 | 1.10 | 1.12 | 1.01 |
| 8 | 0.88 | 0.92 | 0.97 | 0.95 | 23 | 1.00 | 1.08 | 1.08 | 1.08 |
| 9 | 0.94 | 1.01 | 1.05 | 0.98 | 24 | 0.95 | 1.02 | 1.04 | 0.97 |
| 10 | 0.96 | 1.04 | 1.07 | 0.98 | 25 | 0.98 | 0.95 | 0.98 | 0.95 |
| 11 | 0.94 | 1.01 | 1.05 | 0.98 | 26 | 0.96 | 0.93 | 0.96 | 0.93 |
| 12 | 0.96 | 1.03 | 1.05 | 1.03 | 27 | 1.00 | 0.92 | 0.95 | 0.95 |
| 13 | 1.00 | 1.06 | 1.07 | 1.07 | 28 | 0.85 | 0.87 | 0.93 | 0.90 |
| 14 | 0.95 | 1.02 | 1.04 | 1.02 | 29 | 1.12 | 1.05 | 1.08 | 1.08 |
| 15 | 0.95 | 1.02 | 1.04 | 1.02 | 30 | 1.13 | 1.06 | 1.09 | 1.08 |

## D. Case Study Four: IEEE Case 57

This system has 7 generators which are totally generating 1278.66 MW and 321.08 MVAr. There are 42 load buses which are consuming 1250.8 MW and 336.4 MVAr. In this case study, we have added 320 MVAr and 200 MVAr on buses 13 and 55 respectively. As a result, multiple buses including 13, 31-34, 52-55 will go out of their limits. The algorithm will control bus 34 in the first iteration and the result is shown in table V. For the second iteration, we face a two-sided conflict among bus 51 and buses 54, 55. The controlled bus will be chosen as bus 51 and bus 54 which will resolve the voltage violation of bus 51. For the last iteration, bus 54 will be chosen to bring all the voltages back to their normal limits.

TABLE V. Bus voltages in each iteration
(*IEEE* – 57bus)

| Bus | Voltage (pu) | | | | Bus | Voltage (pu) | | | |
|---|---|---|---|---|---|---|---|---|---|
| | *Dist.* | *Iter. 1* | *Iter. 2* | *Iter.3* | | *Dist.* | *Iter. 1* | *Iter. 2* | *Iter.3* |
| 1 | 1.04 | 1.08 | 1.07 | 1.07 | 30 | 0.9 | 0.99 | 0.96 | 0.97 |
| 2 | 1.01 | 1.01 | 1.01 | 1.01 | 31 | 0.87 | 0.97 | 0.93 | 0.94 |
| 3 | 0.99 | 1.06 | 1.05 | 1.05 | 32 | 0.89 | 0.99 | 0.94 | 0.95 |
| 4 | 0.98 | 1.04 | 1.04 | 1.04 | 33 | 0.88 | 0.98 | 0.94 | 0.95 |
| 5 | 0.98 | 1.01 | 1.03 | 1.04 | 34 | 0.897 | 1 | 0.94 | 0.96 |
| 6 | 0.98 | 1 | 1.03 | 1.04 | 35 | 0.91 | 1 | 0.95 | 0.97 |
| 7 | 0.97 | 1 | 1.02 | 1.04 | 36 | 0.92 | 1.01 | 0.96 | 0.97 |
| 8 | 1.01 | 1.02 | 1.06 | 1.07 | 37 | 0.93 | 1.02 | 0.97 | 0.98 |
| 9 | 0.98 | 1.07 | 1.07 | 1.1 | 38 | 0.96 | 1.05 | 1 | 1.01 |
| 10 | 0.98 | 1.06 | 0.99 | 1 | 39 | 0.92 | 1.02 | 0.97 | 0.98 |
| 11 | 0.93 | 1.02 | 0.98 | 1 | 40 | 0.91 | 1.01 | 0.96 | 0.97 |
| 12 | 1.02 | 1.1 | 0.96 | 0.96 | 41 | 0.95 | 1.04 | 1 | 1.02 |
| 13 | 0.896 | 0.99 | 0.92 | 0.93 | 42 | 0.91 | 1.01 | 0.97 | 0.99 |
| 14 | 0.91 | 0.99 | 0.94 | 0.95 | 43 | 0.96 | 1.06 | 1.02 | 1.04 |
| 15 | 0.95 | 1.03 | 0.99 | 1 | 44 | 0.96 | 1.05 | 1 | 1.02 |
| 16 | 1.01 | 1.09 | 0.98 | 0.98 | 45 | 0.99 | 1.08 | 1.03 | 1.04 |
| 17 | 1.02 | 1.08 | 1.02 | 1.02 | 46 | 1 | 1.09 | 1.03 | 1.04 |
| 18 | 0.99 | 1.06 | 1.06 | 1.06 | 47 | 0.97 | 1.07 | 1.01 | 1.02 |
| 19 | 0.95 | 1.02 | 1 | 1.01 | 48 | 0.97 | 1.06 | 1 | 1.02 |
| 20 | 0.93 | 1.01 | 0.98 | 0.99 | 49 | 0.97 | 1.07 | 1.01 | 1.02 |
| 21 | 0.95 | 1.05 | 1 | 1.01 | 50 | 0.98 | 1.07 | 1 | 1.02 |
| 22 | 0.95 | 1.04 | 0.99 | 1.01 | 51 | 1.03 | 1.13 | 1.05 | 1.06 |
| 23 | 0.95 | 1.04 | 0.99 | 1.01 | 52 | 0.87 | 0.94 | 0.96 | 0.98 |
| 24 | 0.95 | 1.02 | 1 | 1.01 | 53 | 0.82 | 0.9 | 0.92 | 0.95 |
| 25 | 0.92 | 1.01 | 0.98 | 0.99 | 54 | 0.76 | 0.87 | 0.89 | 0.92 |
| 26 | 0.91 | 0.98 | 0.96 | 0.97 | 55 | 0.72 | 0.86 | 0.87 | 0.91 |
| 27 | 0.94 | 0.99 | 1 | 1.01 | 56 | 0.91 | 1.01 | 0.97 | 0.99 |
| 28 | 0.96 | 1 | 1.02 | 1.03 | 57 | 0.91 | 1.01 | 0.96 | 0.98 |
| 29 | 0.97 | 1.01 | 1.04 | 1.05 | | | | | |

## V. CONCLUSIONS

In this paper, we have proposed an optimal algorithm to control the voltages of power systems based on the sensitivity matrix. This algorithm finds the best set of inputs by using the eigen decomposition technique. This input will be used to change the generator voltages to reach the desired output for the control bus(es). In case of any conflicts between the nodes, we have modified this set of inputs to clear the conflict. The results achieved from 4 IEEE cases, have shown that the violated buses have been controlled appropriately with the proposed algorithm.

## VI. APPENDIX

**Proof**: let N be a symmetric matrix, then it has three characteristics: 1- It has only real eigenvalues. 2- It has independent eigenvectors for distinct eigenvalues (i.e. $V^{-1}$ $exists$ for the subset of independent eigenvalues and it is equal to $V^T$ ). 3- It has orthogonal eigenvectors for distinct eigenvalues. (i.e. $v_i^T v_j = 0$ for $i \neq j$). By using characteristics number 2 and 3, we can rewrite eigen decomposition of matrix N as in (Ap.1):

$$N = V.\Lambda.V^{-1} = V.\Lambda.V^T \quad (Ap.1)$$

where $\Lambda$ is the diagonal of eigenvalues of matrix N. Therefore, by substituting $X = v_{\lambda max}$ in equation (5), the output will move towards a direction that gives the maximum value which is equal to $\lambda_{max}$:

$$J_{max} = v_{\lambda max}{}^T V.\Lambda.V^T v_{\lambda max} = \lambda_{max}$$